\documentclass[preprint]{elsarticle}

\usepackage{lineno,hyperref}
\pdfoutput=1
\usepackage{color}
\usepackage[normalem]{ulem}
\usepackage{chemfig}
\usepackage[version=3]{mhchem}

\journal{Cryogenics}

\begin{document}

\begin{frontmatter}

\title{Superconducting zinc heat switch for continuous nuclear demagnetization refrigerator and sub-mK experiments}

\author[crc]{Ryo Toda\corref{mycorrespondingauthor}}
\cortext[mycorrespondingauthor]{Corresponding author}
\ead{toda.ryo@mail.u-tokyo.ac.jp}
\author[crc]{Shohei Takimoto}
\author[sphys]{Yuma Uematsu}
\author[crc]{Satoshi Murakawa}
\author[crc,sphys]{Hiroshi Fukuyama\corref{mycorrespondingauthor}}
\ead{hiroshi@kelvin.phys.s.u-tokyo.ac.jp}

\address[crc]{Cryogenic Research Center, The University of Tokyo, 2-11-16, Yayoi, Bunkyo-ku, Tokyo 133-0032, Japan}
\address[sphys]{Department of Physics, The University of Tokyo, 7-3-1, Hongo, Bunkyo-ku, Tokyo 133-0033, Japan}

\begin{abstract}
We have developed and tested a zinc superconducting heat switch suitable for magnetic refrigeration and calorimetric experiments at sub-millikevin (sub-mK) temperatures.
The specific application here is an adiabatic demagnetization refrigerator with two PrNi$_5$ nuclear stages, which can keep a temperature of 0.8~mK continuously, (CNDR) proposed by Toda {\it et} {\it al}. (J. Phys.: Conf. Ser. {\bf 969}, 012093 (2018).  
The switch consists of six high-purity zinc foils of 0.25~mm thick which contact seven silver foils by diffusive bonding.
The silver foils are electron beam welded to silver rods that are thermal links to other components. 
The choice of the thin zinc foils is due to reduce the magnetic latent heat  on switching and the contact thermal resistance under a constraint on the aspect ratio of the switch element.
The measured thermal conductance of the whole switch assembly in the normal (closed) state, $K_\mathrm{closed}$, agrees very well down to 70~mK with the value estimated from the residual electrical resistance 114~n$\mathrm{\Omega}$ at 4.2~K, indicating the validity of the Wiedemann-Franz law for zinc.
The measured thermal conductance in the superconducting (open) state, $K_\mathrm{open}$, follows nicely the prediction from the BCS theory and approaches the value expected from the Debye model for thermal phonons near 70~mK.
The heat leak through the HSW from the higher temperature side of 30~mK at most is estimated to be less than 0.5~nW, which is much lower than the expected cooling power ($= 10$~nW) of the CNDR at 0.8~mK .
The switching ratio $K_\mathrm{closed}/K_\mathrm{open}$ extrapolated to 30~mK, is as high as 5$\times10^4$.
All the test results meet the requirements for the CNDR and, for example, heat capacity measurements at sub-mK.
\end{abstract}


\end{frontmatter}


\section{Introduction}
\label{intro}

An ultra-low temperature environment below 1~millikelvin (mK) achievable with nuclear refrigeration has been monopolized by researchers in limited fields such as quantum fluids and solids for many years since its first realization in the 1970s~\cite{Pobell}.
However, recently, sub-mK temperatures is attracting more attention in growing numbers of fields such as material science~\cite{Clark2010}, cryogenic particle detector~\cite{Shirron2006}, nano/micro-mechanical devices~\cite{Palma2017}, and so on.
In such an extreme condition, where thermal fluctuations are substantially reduced, new phenomena and functions emerge in  materials and devices.

In 2018, Toda \textit{et al}.~\cite{Toda2018} proposed a first realistic design of compact and continuous nuclear-spin demagnetization refrigerator (CNDR) which can keep the sample temperature at 0.8~mK continuously with a cooling power of more than 10~nW based on numerical simulations.
The CNDR contains two independent demagnetization stages of PrNi$_5$, a hyperfine enhanced nuclear spin system~\cite{Pobell}, connecting the sample stage and the mixing chamber (MC) of a dilution refrigerator (DR) in series with two heat switches.
Once it becomes available, researches, who are not familiar with ultra-low temperature techniques, can access more easily to the sub-mK environment. 
In addition, this new refrigerator is so compact that it can conveniently be installed on the MC of existing DRs.
After the proposal of Ref.~\cite{Toda2018}, the development of CNDR is now being extensively conducted~\cite{Schmoranzer2019,Takimoto2020,Schmoranzer2020,Takimoto-new}.

To realize the CNDR, it is crucial to develop a high performance heat switch.
Requirements for the relevant heat switch are rather severe compared to those for existing heat switches.
For example: 
\begin{itemize}
\item The maximum allowable thermal residence in the closed (conductive) mode should be of the order of 100~n$\rm{\Omega}$ in the unit of electrical resistance ~\cite{Toda2018,Schmoranzer2019}. 
\item The heat leak through the HSW in the open (non-conductive) mode should be less than 1~nW.
\item The heat generation upon switching should be as small as possible.
\item The whole assembly should be as compact as possible, etc.
\end{itemize}
Previously, many heat switches usable in mK to sub-mK regions have been developed, and all of them are superconducting heat switches making use of a large difference in thermal conductivity between the normal (conductive) state and the superconducting (non-conductive) state~\cite{Pobell}.
However, non of them meets all the requirements for the CNDR application.

Here we report design and construction details of a compact and high-performance superconducting heat switch made of zinc (Zn), which is usable for the CNDR.
Measurement results of its thermal conductances in the closed and open modes down to 70~mK are also given. 
Not only for adiabatic demagnetization refrigeration, heat switch is also a crucial part for calorimetric experiments such as heat capacity (HC) measurements, an important tool in material science.
In this article, through the development of the heat switch for the CNDR, we will also provide useful information on designing heat switches for HC measurements at sub-mK temperatures.

\section{Design}
\label{sec:2}

\begin{figure}[h]
\centering
\includegraphics[width=0.75\linewidth]{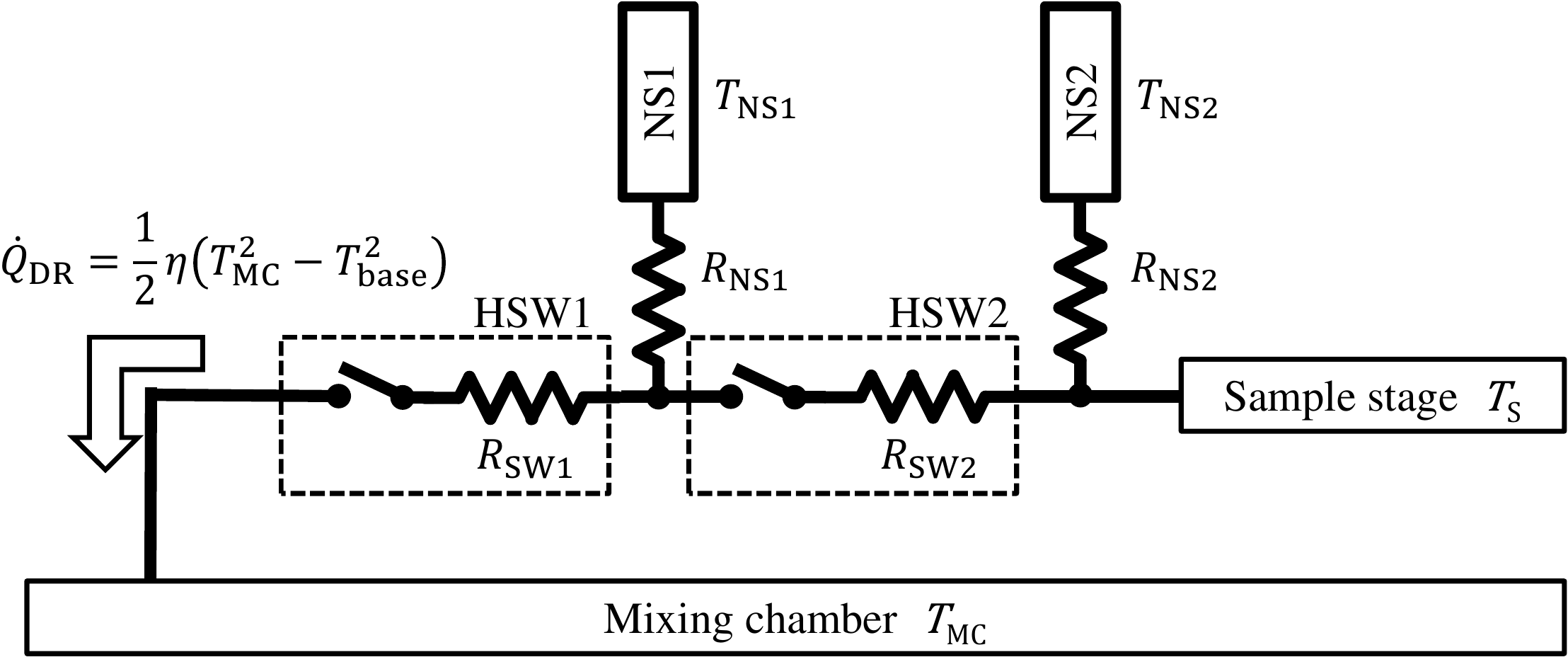}
\caption{Schematic diagram of the CNDR showing a network of various thermal resistances. Here, $R$ is in the unit of electrical resistance. Two nuclear demagnetization stages (NS1 and NS2) are connected in series between the mixing chamber of a dilution refrigerator and the sample stage through two heat switches (HSW1 and HSW2). See Ref.~\cite{Toda2018} for further details.}
\label{fig:schematics}
\end{figure}

Fig.~\ref{fig:schematics} shows a schematic diagram of the CNDR, where we need two heat switches (HSW1 and HSW2).
HSW1 is placed between the MC of a DR and the first nuclear stage (NS1), working just like an ordinary HSW for adiabatic demagnetization refrigerator. 
HSW2 is placed between NS1 and the second nuclear stage (NS2) which is directly attached to the sample stage.
The function of NS2 is to keep the sample stage temperature constant absorbing an ambient heat leak and a heat generated by measurement with slow and continuous demagnetization.
HSW2 is set to the open  mode during precooling and demagnetization cooling of NS1, while to the closed  mode during entropy pumping from NS2 (see Ref.~\cite{Toda2018} for further details).
The essential configuration of the CNDR is nearly the same as that of the continuous adiabatic demagnetization refrigerator (ADR) using paramagnetic salts as a magnetic coolant which now reaches $T =50$~mK for the lowest sustainable temperature~\cite{Shirron2016}.
However, because the operation temperature of the CNDR is more than sixty times lower than ADR, completely different cryogenic techniques should be applied, particularly, to heat switch.

\subsection{Basic design}

In general, there are several requirements for heat switches such as high switching ratio of the thermal conductance in the closed mode ($K_\mathrm{closed}$) to that in the open mode ($K_\mathrm{open}$), rapid and easy switching, small heat generation by switching, etc.
Thermal conductance between components of the CNDR directly affects cooling performance of it~\cite{Toda2018}. Then, for our purose, the thermal conductance in the closed mode ($K_\mathrm{closed}$) of the HSW should be as high as those of other parts made of highly conducting annealed pure metals.  For HC measurements, highly thermal conductance is also favored to reach and measure faster and lower temperatures.
To meet this requirement, superconducting heat switch of a pure metal  is currently the best choice.

First of all, to select a proper superconducting material, it is useful to start to consider with  $K_\mathrm{open}$.
Generally, the thermal conductance $K_{i}$ by carriers $i$ (= phonons (ph) or electrons (el)) is given by 
\begin{equation}
\label{eq: eq1}
K_{i} = \frac{1}{3}\alpha c_{i}v_{i} \lambda_{i},
\end{equation}
where $\alpha$ is the aspect ratio, i.e., the cross section divided by the length of the element, and $v_{i}$ and $\lambda_{i}$ are the mean velocity and the mean free path of the carriers, respectively.
At low temperatures of $T \leq 0.1T_\mathrm{c}$ where only few unpaired normal electrons remain, major thermal carriers are phonons, and the specific heat of the phonons is expected to be $c_{\mathrm{ph}} \propto (T/\theta_{\rm{D}})^3$ from the Debye model.
Here, $T_\mathrm{c}$ is the superconducting transition temperature, and $\theta_{\rm{D}}$ is the Debye temperature of the element.
In the Debye model, $v_{\mathrm{ph}}$ is propotional to the $\theta_{\rm{D}}$. 
Thus, from Eq.~\ref{eq: eq1}, we have $K_\mathrm{open} \propto (T^{3}/\theta_{\rm{D}}^{2})$.
Note that we can safely assume that $\lambda_{\mathrm{ph}}$ is temperature independent at $T \leq 30$~mK for superconducting elements with minimum dimensions less than 1~mm~\cite{Gloos1990}, and this is our case (see later).
So, $K_\mathrm{open}$ at $T \leq 0.1T_\mathrm{c}$ is determined mainly by $T_\mathrm{c}$ as well as $\theta_{\rm{D}}$ of the metal element.
On the other hand, at $T \geq 0.1T_\mathrm{c}$, $K_\mathrm{open}$ increases exponentially due to increasing population of the unpaired electrons up to $T = T_\mathrm{c}$, which makes difficult for the HSW to keep good thermal insulation. 
Therefore, $T_\mathrm{c}$ must be at least ten times higher than the highest operation temperature ($T_\mathrm{highest}$) of the HSW.
We assume that $T_\mathrm{highest} = 15$~mK for HSW1 and 30~mK for HSW2, thus the relevant HSW element must have $T_\mathrm{c} \geq 150$ and 300~mK, respectively.
Note that, in this article, we set $T_\mathrm{highest} = 30$~mK for HSW2 rather than 22~mK, the simulation result in Ref.~\cite{Toda2018}, for safety design.

The other consideration on the material choice is the magnetic latent heat, $Q_\mathrm{mag} = (1/2)\mu_0\{H_\mathrm{c}(T)\}^2$, released when the HSW is turned off by decreasing the applied magnetic field $H$ below the superconducting critical field $H_\mathrm{c}$.
Here, $\mu_0$ is vacuum permeability. 
Larger $Q_\mathrm{mag}$ is unfavorable, because it causes sample temperature instability when switching in the case of CNDR and rises the lowest temperature of HC measurements.
In type-I superconductors, $H_\mathrm{c}$ is roughly proportional to $T_\mathrm{c}$.
Thus, $T_\mathrm{c}$ should not be too high compared to $10T_\mathrm{highest}$ to keep the heat generation as small as possible.

Table~\ref{SC} shows various physical properties of candidate pure metals for the HSW element~\cite{Pobell}.
From the above mentioned considerations, we have chosen zinc ($T_\mathrm{c} =$~850~mK, $H_\mathrm{c} =$~5.4~mT) rather than  aluminum ($T_\mathrm{c} = 1,180$~mK, $H_\mathrm{c} = 10.5$~mT) due to the lower $H_\mathrm{c}$, i.e., smaller $Q_\mathrm{mag}$. 
It is noted that, for applications where the switching ratio is more important than the switching heat, aluminum could be a better choice.
So far, only a few Zn HSWs have been developed~\cite{Krusius1978,Tajima2003}.
One of them has too low $K_\mathrm{closed}$~\cite{Krusius1978}, and the other is less compact~\cite{Tajima2003} for the present purpose.
In the latter work, $K_\mathrm{closed}$ is not known.

\begin{table}[h]
\begin{center}
\caption{Properties of some superconducting pure metals~\cite{CRCBook, Stewart1983}. $T_\mathrm{m}$ is the melting temperature, and $\rho_\mathrm{RT}$ is the electrical resistivity at $T = 273$~K. See text for other symbols.}
\label{SC}
\vspace{2mm}
\begin{tabular}{lccccc}
\hline \hline
material & $T_\mathrm{c}$ & $H_\mathrm{c}$ & $\theta_\mathrm{D}$ & $T_\mathrm{m}$ & $\rho_\mathrm{RT}$ \\
 & (K)  &  (mT)  &  (K)  & ($^\circ$C)  & ($\mu\mathrm{\Omega}\cdot$cm) \\ \hline
Pb& 7.20 & 80.3 & 105 & ~327 & 19.2 \\
Sn& 3.72 & 30.5 & 199 & ~232 & 11.5 \\
In& 3.41 & 28.2 & 112 & ~157 &  8.0 \\
Al& 1.18 & 10.5 & 433 & ~660 &  2.4 \\
Zn& 0.85 & ~5.4 & 329 & ~420 &  5.5 \\
Cd& 0.52 & ~2.8 & 210 & ~321 &  6.8 \\
Ti& 0.40 & ~5.6 & 420 & 1670 & 39.0 \\
\hline \hline
\end{tabular}
\end{center}
\end{table}

\subsection{Constraint for aspect ratio of superconducting element}

Next, we consider constraints on the aspect ratio $\alpha$ of the Zn element.
Since a Zn region of $K_\mathrm{open}$ and $K_\mathrm{closed}$ are proportional to $\alpha$, they provide a maximum allowable aspect ratio $\alpha_\mathrm{max}$ and a minimum allowable one $\alpha_\mathrm{min}$, respectively.
The heat leak $\dot{Q}_\mathrm{through}$ in the open mode from the high temperature side at $T = T_\mathrm{high}$ through the HSW is mainly determined by $T_\mathrm{high}$ not by the temperature of the cold side $T_\mathrm{low}$, as it is proportional to $\alpha(T_\mathrm{high}^4 - T_\mathrm{low}^4)$.
For HSW1 in the CNDR, $T_\mathrm{high}$ will be the same as the base temperature of DR with no heat load ($T_\mathrm{base} \approx 15$~mK) during demagnetization of NS1.
For HSW2, it can be as high as 30~mK during precooling of NS1.
In order to reduce $\dot{Q}_\mathrm{through}$ less than 0.5~nW, one-twentieth of expected cooling power of CNDR ($\approx 10$~nW), $\alpha_\mathrm{max}$ for HSW1 and HSW2 should be 29 and 1.8~mm, respectively, for example when $\lambda_{\mathrm{ph}} = 0.25$~mm (see the next section).
Apparently, this constrain is severer for HSW2 than for HSW1.

On one hand, higher $K_\mathrm{closed}$ is crucial in determining the cooling power of the CNDR which determines the lowest operation temperature under a constant heat leak~\cite{Toda2018,Schmoranzer2019}, and the lowest measurable temperature in case of HC.
In the followings, we discuss residual electrical resistances $R$ at 4.2~K, which are measurable much more easily, instead of $K_\mathrm{closed}$.
$R$ can be converted to $K_\mathrm{closed}$ and vice versa through the Wiedemann-Franz law:
\begin{equation}
\label{eq: WFlaw}
K_\mathrm{closed} = LT/R,
\end{equation}
where $L$ is the Lorenz number ($= 2.44\times 10^{-8}$~W$\rm{\Omega}$K$^{-2}$).
In the following estimations, we used measured RRR values obtained after proper annealing for the Zn (RRR = 2,000) and Ag (RRR = 3,000), a thermal link for HSWs (see the next section for further details).
Here RRR is the residual resistivity ratio.

Let us consider $K_\mathrm{closed}$ of HSW2 first.
The spin entropy should conduct from NS2 through the series resistance, $R_\mathrm{NS2} + R_\mathrm{HSW2} + R_\mathrm{NS1}$, as efficiently as possible (see Fig.~\ref{fig:schematics}).
Here, $R_\mathrm{NS1}$, $R_\mathrm{NS2}$ and $R_\mathrm{HSW2}$ are the residual electrical resistances of NS1, NS2 and HSW2 in the closed mode, respectively.
Since a realistic evaluation for $R_\mathrm{NS2}$ or $R_\mathrm{NS1}$ is quite low ($\approx 140$~n$\mathrm{\Omega}$)~\cite{Takimoto-new}, $R_\mathrm{HSW2}$ should also be of the order of $140$~n$\mathrm{\Omega}$ or less not to spoil the cooling power of the CNDR. 
This means $\alpha_\mathrm{min} = 0.21$~mm if the main resistance of HSW2 is in the Zn region.
After all, combing with $\alpha_\mathrm{max}$ obtained from $K_\mathrm{open}$ earlier, we have a constraint $0.21 < \alpha < 1.8$~mm for HSW2.

In the case of HSW1, the constraint on $K_\mathrm{closed}$ is much less tight, because the conduction efficiency is usually determined by the cooling power of the DR than thermal conductivity of the heat links made of pure metals.

The cooling power of the DR is known to obey approximately the relation $\dot{Q}= \eta (T_\mathrm{MC}^2-T_\mathrm{base}^2)$, where the coefficient $\eta$ is predominantly determined by the $^3$He circulation rate. This temperature dependence is the same as the heat conduction of metals $\dot{Q}= \frac{L}{2 R} (T_\mathrm{high}^2-T_\mathrm{low}^2)$. 
Then the \textit{effective} (electrical) resistance of DR can be considered as $R_\mathrm{DR} = L / (2 \eta)$.
In a standard DR, $\eta = 0.01$~W/K$^2$, i.e., $\dot{Q} = 100$~$\mathrm{\mu}$W at 100~mK, which corresponds to $R_\mathrm{DR} = 1.2$~$\mathrm{\mu\Omega}$.
As the precooling time of NS1 in the CNDR is determined by the series resistance, $R_\mathrm{NS1} + R_\mathrm{HSW1} + R_\mathrm{DR}$ (see Fig.~\ref{fig:schematics}), $R_\mathrm{HSW1}$ should be smaller than $R_\mathrm{DR}$ not to spoil the cooling power of the DR, hence the cooling power of the CNDR.
Here, $R_\mathrm{HSW1}$ is the residual electrical resistances of HSW1 in the closed mode.
Note that the \textit{effective} cooling power of CNDR can be considered as a total heat which can be absorbed from the sample stage divided by a total time for the precooling--demagnetization--entropy-pumping--remagnetization cycle~\cite{Toda2018}.
This gives $\alpha_\mathrm{min} = 0.025$~mm.
Finally, combing with $\alpha_\mathrm{max}$ obtained from $K_\mathrm{open}$ earlier, we have a constraint $0.025 < \alpha < 29$~mm for HSW1, which is much wider than for the HSW2.
Consequently, in the following fine tuning of design (specific sizes or shapes), we will concentrate on that of  HSW2, since the resultant optimized condition will automatically meet that for HSW1.

%
%

\subsection{Optimization of dimensions}
\label{sec:optimization}

For the compactness of HSW, we restricted ourselves to accommodate the HSW element within a coil bobbin of a small superconducting solenoid for switching with the inner diameter of 10~mm (outer diameter is 22~mm).
This inevitably rules out to use screw-fastened demountable joint which is rather large in size~\cite{Mueller1990}.
It is also known that soft soldering does not provide an acceptably low contact resistance presumably due to a rather thick contact alloy layer~\cite{Meijer1974}.
Then, we adopted to contact thin Zn foils, which is the switch element, and thin Ag foils, which is the thermal link, with the diffusive bonding technique~\cite{Tajima2003,Bunkov1989}.
There are several benefits of using the thin foils.
The first one is to increase the contact area substantially.
The second one is to retain the mechanical flexibility of the switch element which protects the contact from a possible large stress force caused by differential thermal contractions among various construction materials including thermal isolation supports made by Vespel SP-22 in our case.
The third benefit is that, by decreasing $d$ but keeping $\alpha$ constant, we can decrease $K_{\mathrm{open}}$ hence increase the switching ratio.
This is because $K_{\mathrm{open}}$ is proportional to $\lambda_\mathrm{ph}$ (see Eq.(\ref{eq: eq1})), and the intrinsic $\lambda_\mathrm{ph}(T)$ ($\propto 1/T$) can be limited by $d$ at low temperatures~\cite{Gloos1990}.
However, there are practical limits for the thinning.
One is easy handling.
The other is that the thermal conductivity of pure metals tends to decrease if $d$ is reduced below 0.1~mm keeping $\alpha$ fixed, which decreases the switching ratio.
This is due to diffusive scatterings for conduction electrons at surfaces~\cite{Gloos1990}.
Consequently, we decided to use a Zn foil of 0.25~mm thick and an Ag foil of 0.5~mm thick.

\begin{figure}[h]
\centering
  \includegraphics[width=1.00\linewidth]{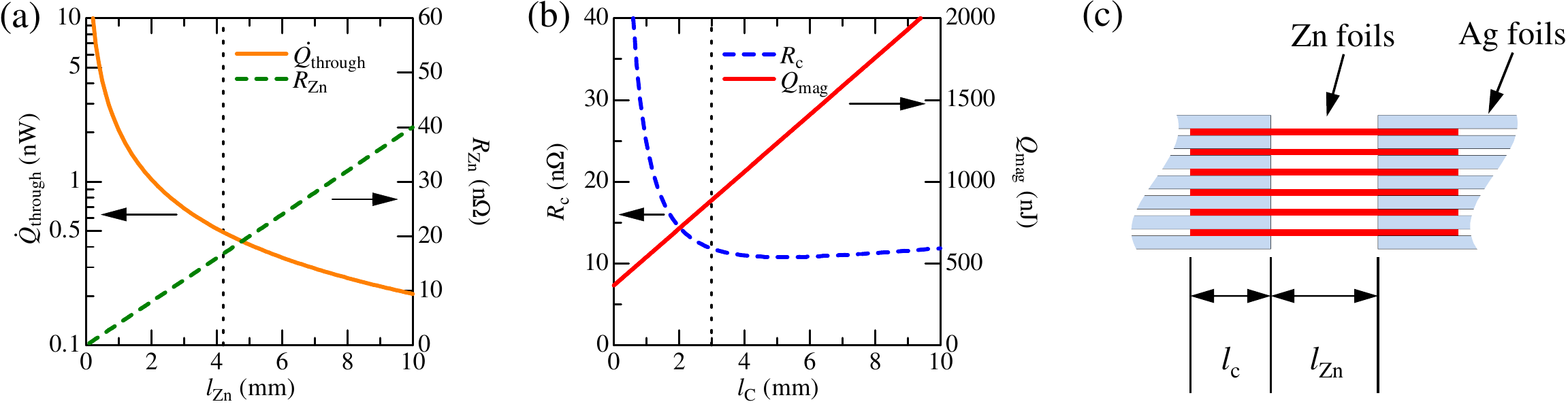}
   \caption{(a) Calculated heat leak through the HSW ($\dot{Q}_\mathrm{through}$; solid line) and electrical resistivity of the zinc element ($R_\mathrm{Zn}$; dashed line) in the open mode as functions of the length of zinc element ($l_\mathrm{Zn}$). (b) Calculated contact resistance ($R_\mathrm{c}$; dashed line) and magnetic latent heat ($Q_{\mathrm{mag}}$; solid line) as functions of the Zn/Ag contact length $l_\mathrm{c}$. (c) Configuration of the central part of the HSW.}
\label{fig:design}
\end{figure}

To utilize the bore space of the coil bobbin effectively, we sandwiched six Zn foils of 5~mm wide  with seven Ag foils of the same width and diffusively bonded them together.
At this moment, there still remain two adjustable parameters, i.e., the length of bare Zn part ($l_\mathrm{Zn}$) and the contact length ($l_\mathrm{c}$) of the Zn/Ag bonding.
As $l_\mathrm{Zn}$ increases, the heat leak $\dot{Q}_\mathrm{through}$ in the open mode can be reduced as shown by the solid (orange) line in Fig.~\ref{fig:design}(a). 
Whereas, as explained in the previous sections,  increasing $l_\mathrm{Zn}$ results in increasing the closed-state resistance $R_\mathrm{Zn}$ as shown by the dashed (green) line in the figure.
A good compromise is to set $l_\mathrm{Zn} = 4.2$~mm, with which $\dot{Q}_\mathrm{through} =$~0.49~nW and $R_\mathrm{Zn} =$~17~n$\mathrm{\Omega}$.
The dimensions correspond to $\alpha = 1.8$, the highest bound of the constraint for HSW2 ($0.21 < \alpha < 1.8$).
Remember again that, in this article, we assumed a higher $T_\mathrm{highest}$ (= 30~mK) than the simulated result (= 22~mK)~\cite{Toda2018} for safety design, the actual $\dot{Q}_\mathrm{through}$ would be less than 0.2~nW.

Finally, we would like to decide $l_\mathrm{c}$.
In general, the total resistance ($R_\mathrm{c}$) of a joint contact of two parallel plates, which contact each other over a length $l_\mathrm{c}$, is analytically given in Ref.~\cite{Ishimoto1989} in terms of bulk resistivities of the two plates and contact resistivity between the plates.
A calculated result of $R_\mathrm{c}$ for our HSW is shown as a function of $l_\mathrm{c}$ as the (blue) dashed line in Fig.~\ref{fig:design}(b).
In this calculation, we used the measured resistivities of the Zn and Ag foils ($\rho_\mathrm{Zn} = 30$~p$\mathrm{\Omega}\cdot$m and $\rho_\mathrm{Ag} = 5.3$~p$\mathrm{\Omega}\cdot$m) and the measured contact resistivity between them ($\rho_\mathrm{c} = 14$~n$\mathrm{\Omega}\cdot$cm$^2$) (see the next section for details).
$R_\mathrm{c}$ has an effective length ($\approx 3$~mm) beyond which it practically saturates or even has a shallow minimum near 5~mm.
In the same figure, we also plot a calculated magnetic latent heat ($Q_\mathrm{mag}$), which is released when the switch turns off, as a function of $l_\mathrm{c}$ by the solid (red) straight line.
Note that, in addition to the bare Zn part of 4.2~mm long ($l_\mathrm{Zn}$), the whole contact area of $l_\mathrm{c}$ long should be located in the applied magnetic field for switching. 
Again, we can find a good compromise, $l_\mathrm{c} = 3.0$~mm, with which $2R_\mathrm{c} =$~24~n$\mathrm{\Omega}$ and $Q_\mathrm{mag} =$~890~nJ.
The $2R_\mathrm{c}$ is much smaller than $R_\mathrm{NS1}$ and $R_\mathrm{NS2}$ ($\approx 140$~n$\mathrm{\Omega}$ each) as expected.

\section{Construction of heat switch}
\label{sec:construction}

In this section, we show details of various heat treatments of construction materials of the HSW and measurement results of their residual electrical resistances at $T = 4.2$~K.

\subsection{Heat treatments and residual electrical resistance measurements}

We used a 99.9998\% pure Zn foil (The Nilaco Co.) of 0.25~mm thick for the superconducting element and a pure Ag foil of 0.5~mm thick and rods (Tokuriki Honten Co., Ltd., $>99.99$\%) for the thermal links.
The Zn foil was etched by a mixture of HNO$_3$ (60\%(w/w) : H$_2$O$_2$ (30\%(w/w)) : C$_2$H$_5$OH $= 3:7:10$ in volume ratio.
And then, it was annealed at 200~$^\circ$C for 16 hours being encapsuled in a quartz tube filled with a helium atmosphere of $P = 0.05$~MPa.
This heat treatment increased the RRR value from 1,200 to 2,000 corresponding to the residual resistivity of 30~p$\mathrm{\Omega}\cdot$m with a little mass loss by evapolation.
The Ag foil and rods were etched by a dilute HNO$_3$ water solution and then annealed at 645~$^\circ$C for 3 hours in an air flow ($P = 0.1$--$0.3$~Pa at the location of the materials).
This heat treatment increased the RRR value up to 3,000 for the Ag-film (Tokuriki Honten Co., Ltd., $>99.99$\%) corresponding to the residual resistivity of 5.3~p$\mathrm{\Omega}\cdot$m. While, in the case of the Ag-rods (Tokuriki Honten Co., Ltd., $>99.99$\%), the RRR value increased only to 800, presumably due to difference of the purity in range of $<$0.01\% of the starting material.

We optimized the Zn/Ag diffusive bonding by varying the heat treatment temperature using a test piece with a contact area of $5\times5$~mm$^2$.
A contact pair of annealed Zn foil and Ag foils was sandwiched by two tungsten sheets of 0.1~mm thick and two stainless steel plates of 12~mm thick.
The stuck was pressed under a pressure of $P = 170$~MPa with an oil-piston pressurizer and then clamped by stainless steel screws to keep the pressure.
The clamped whole piece was heated in vacuum for $t = 3$~hours at various temperatures.
The warming from room temperature to the target temperature took about 3~hours, and the cooling back to room temperature took 6 hours or longer.
The tungsten sheets were inserted to protect the Zn/Ag pair from contamination during the heating.

\begin{figure}[h]
\centering
\includegraphics[width=0.45\linewidth]{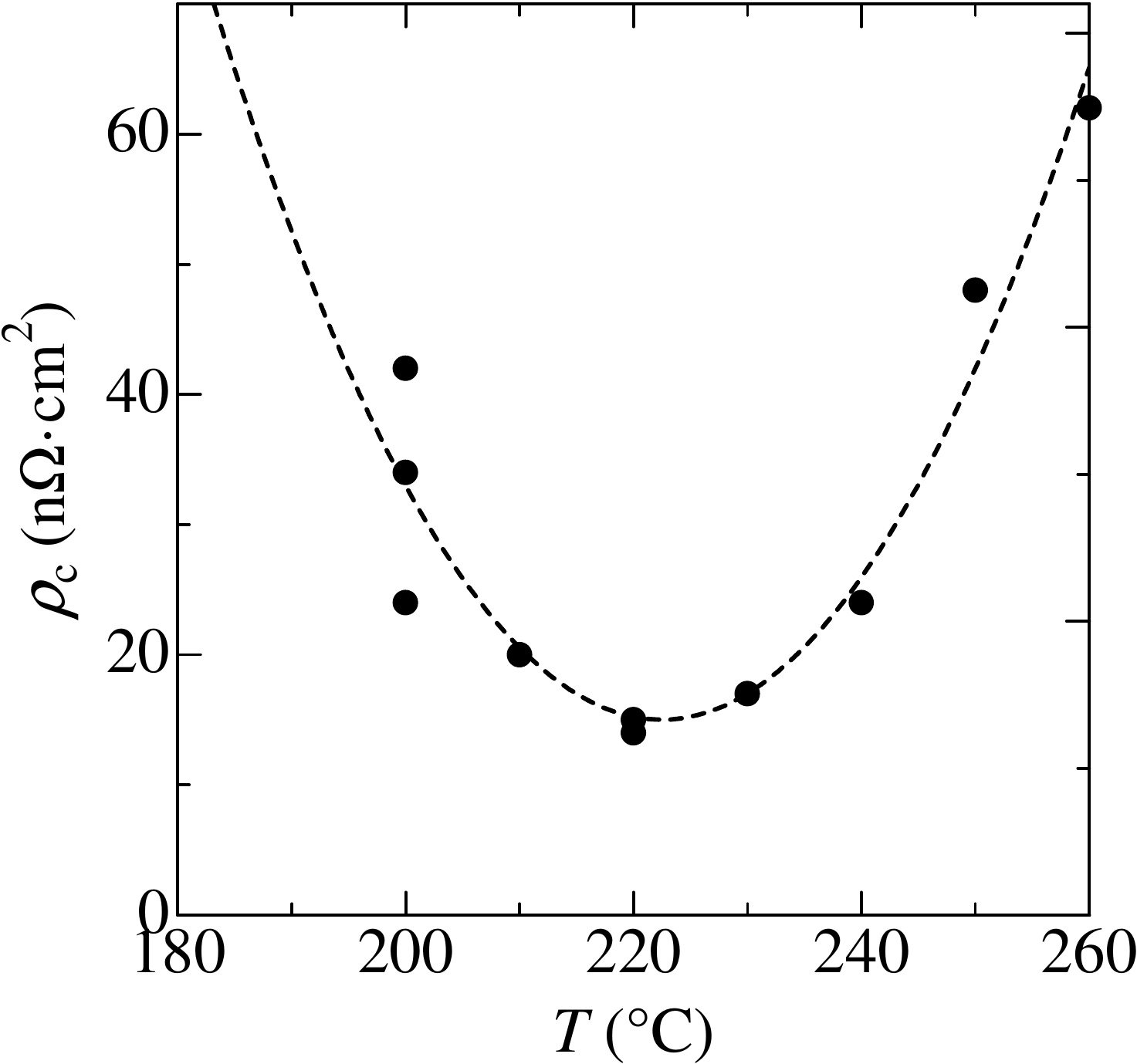}
\caption{Measured contact resistivity $\rho_\mathrm{c}$ of Zn/Ag contacts diffusively bonded at different temperatures. Other heat-treatment parameters such as the heating time (3 hours) and the applied pressure (170~MPa) are fixed. The dashed line is a guide to the eye.}
\label{fig:contactres}       
\end{figure}

Fig.~\ref{fig:contactres} shows measured residual contact resistivities $\rho_\mathrm{c}$ of the test pieces diffusively bonded at different temperatures.
$\rho_\mathrm{c}$ was evaluated from a measured total resistance including the contact region taking account of bulk resistivities of the Zn and Ag foils measured at different positions from the contact region based on the recipe described in the previous section.
The plot clearly shows a minimum of $\rho_\mathrm{c} = 14$~n$\mathrm{\Omega}\cdot$cm$^2$ at $T_\mathrm{op} = 220$~$^\circ$C, an optimum heat treatment temperature.
As $T$ increases above $T_\mathrm{op}$, $\rho_\mathrm{c}$ increases probably because the thickness of an interface layer, where Zn and Ag atoms  interpenetrate, increases.
As $T$ decreases below $T_\mathrm{op}$, the area of the interface layer probably decreases.
If so, we may have  slightly different $T_\mathrm{op}$ values for different $t$.
The optimized $\rho_\mathrm{c}$ ($= 14$~n$\mathrm{\Omega}\cdot$cm$^2$) for the Zn/Ag contact obtained here is almost the same as the previously reported values for the Al/Cu ($\rho_\mathrm{c} =12$~n$\mathrm{\Omega}\cdot$cm$^2$, $T = 500$~$^\circ$C, $t = 15$~min.)~\cite{Bunkov1989} and Zn/Cu (14~n$\mathrm{\Omega}\cdot$cm$^2$, 200~$^\circ$C, 15~min.)~\cite{Tajima2003} contacts obtained by the similar diffusive bonding technique.

\subsection{Assembling}

\begin{figure}[h]
\centering
\includegraphics[width=0.80\linewidth]{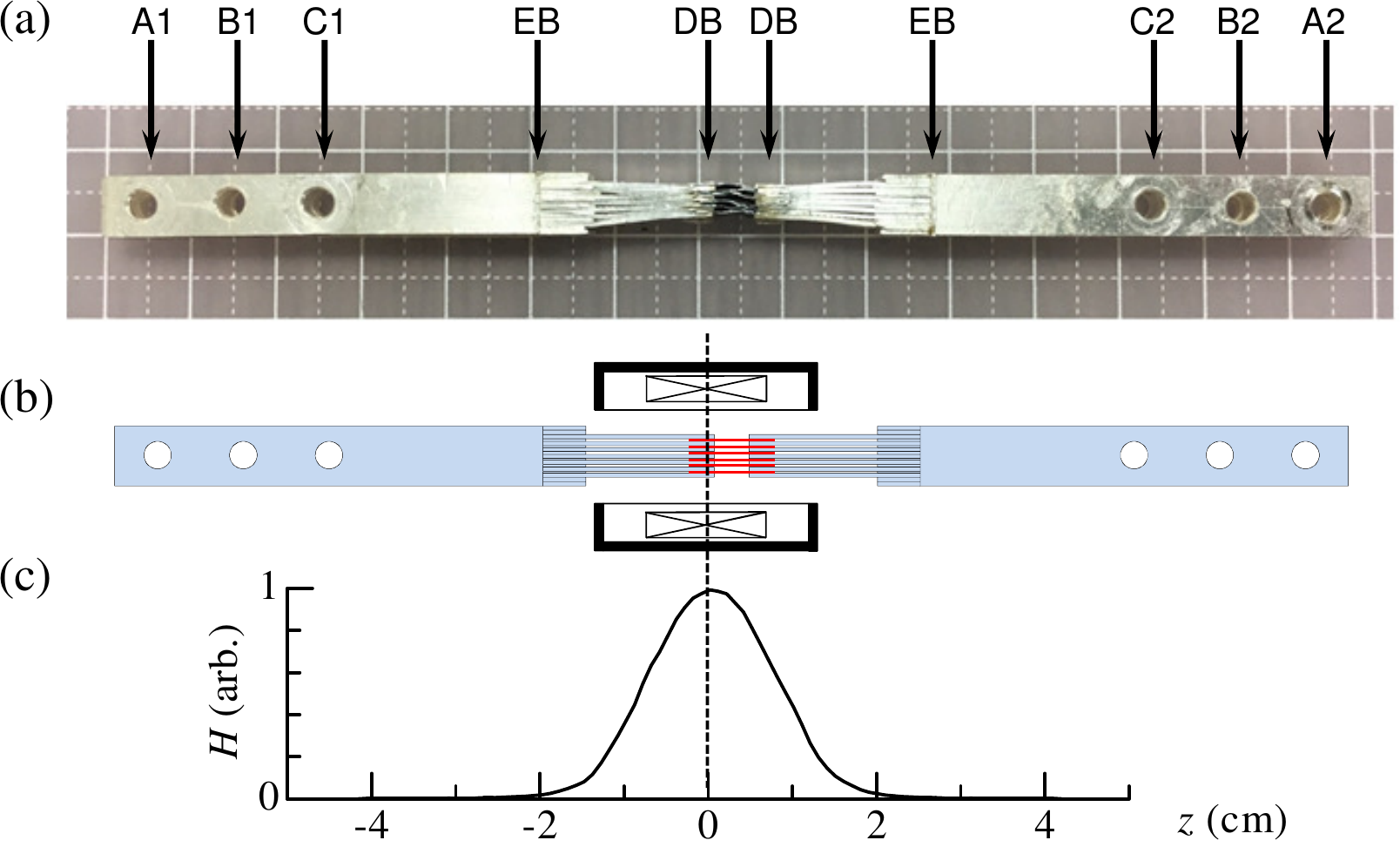}
\caption{(a) Photo and (b) schematic drawing of the assembled HSW. (c) Magnetic field profile $H$ produced by the small superconducting magnet for switching along the central axis of the magnet. Note that the center of the magnet is offset by 3.1~mm from the HSW center to reduce the magnetic latent heat.}
\label{fig:body}       
\end{figure}

Fig.~\ref{fig:body} shows a photo and a schematic drawing of the HSW  constructed following the design we discussed above.
The seven Ag foils were first electron beam welded to an Ag rod of  $5\times7\times50$~mm$^3$ size together with several spacer Ag short foils at the place denoted as ``EB" in the figure.
After the electron beam welding, such a Ag thermal link with seven fins was annealed following the above mentioned recipe.
Two such thermal links were then diffusively bonded with the six Zn foils at the places denoted as ``DB" in the figure under the optimum condition.
The residual total resistance of the assembled HSW ($R_\mathrm{HSW}$) was measured at 4.2~K with the four-terminal method making use of drill holes on the rods.
The measured $R_\mathrm{HSW}$ was 114~n$\mathrm{\Omega}$, meeting our requirement.
It agrees reasonably well with the expected value ($=107$~n$\mathrm{\Omega}$) estimated from the test measurements for individual parts including the known residual resistance of an electron-beam welded Ag piece similar to the present case in size ($= 13.5$~n$\mathrm{\Omega}$~\cite{Takimoto-new}). 

Fig.~\ref{fig:body} also shows the location of the superconducting solenoid, which is offset by 3.1~mm with respect to the center of the Zn foils in order to produce an asymmetric magnetic field distribution so that  $Q{_\mathrm{mag}}$  conducts selectively to the high temperature side of the HSW.
This arrangement is known to reduce a fraction of $Q{_\mathrm{mag}}$, which flows to the low temperature side, down to a ....th~\cite{Matsumoto2003} and to be very effective to measure a tiny heat capacity of 0.1~mK/K down to 0.3~mK~\cite{Sato2010}.

The eddy current heating $\dot{Q}{_\mathrm{eddy}}$ by switching is calculated as below 0.1~nW at most assuming that the external field is swept by 15~mT across $H{_\mathrm{c}}$ in 100~seconds.
Because there are only thin zinc and silver foils in the magnetic field, eddy current heating is suppressed enough. You might imagine that there is a large loss at the contact point of Zn/Ag, but there is the contact resistance between thin films, eddy current heating in contact region is small contrary to appearance.
If this is troublesome, $\dot{Q}{_\mathrm{eddy}}$ can easily be reduced by slowing the sweep rate $\dot{H}$ (cf. $\dot{Q}{_\mathrm{eddy}} \propto \dot{H}^2$).

\section{Thermal conductance measurements}
\label{sec:5}

We installed the HSW assembly onto the MC of a DR and measured its thermal conductances both in the open and closed modes in a temperature range from 70 to 900~mK.
Specifically, by connecting one end of the HSW firmly to the MC with screws at the ``A2" and ``B2" positions and by flowing a known heat current generated by a heater fixed at ``A1" on the other free side, a temperature difference between ``C1" and ``C2" was measured with RuO$_2$ resistance thermometers (nominal 470~$\mathrm{\Omega}$, ALPS Electric Co., Ltd.) fixed at those positions (see Fig.~\ref{fig:body}).

\begin{figure}[h]
\centering
  \includegraphics[width=0.55\linewidth]{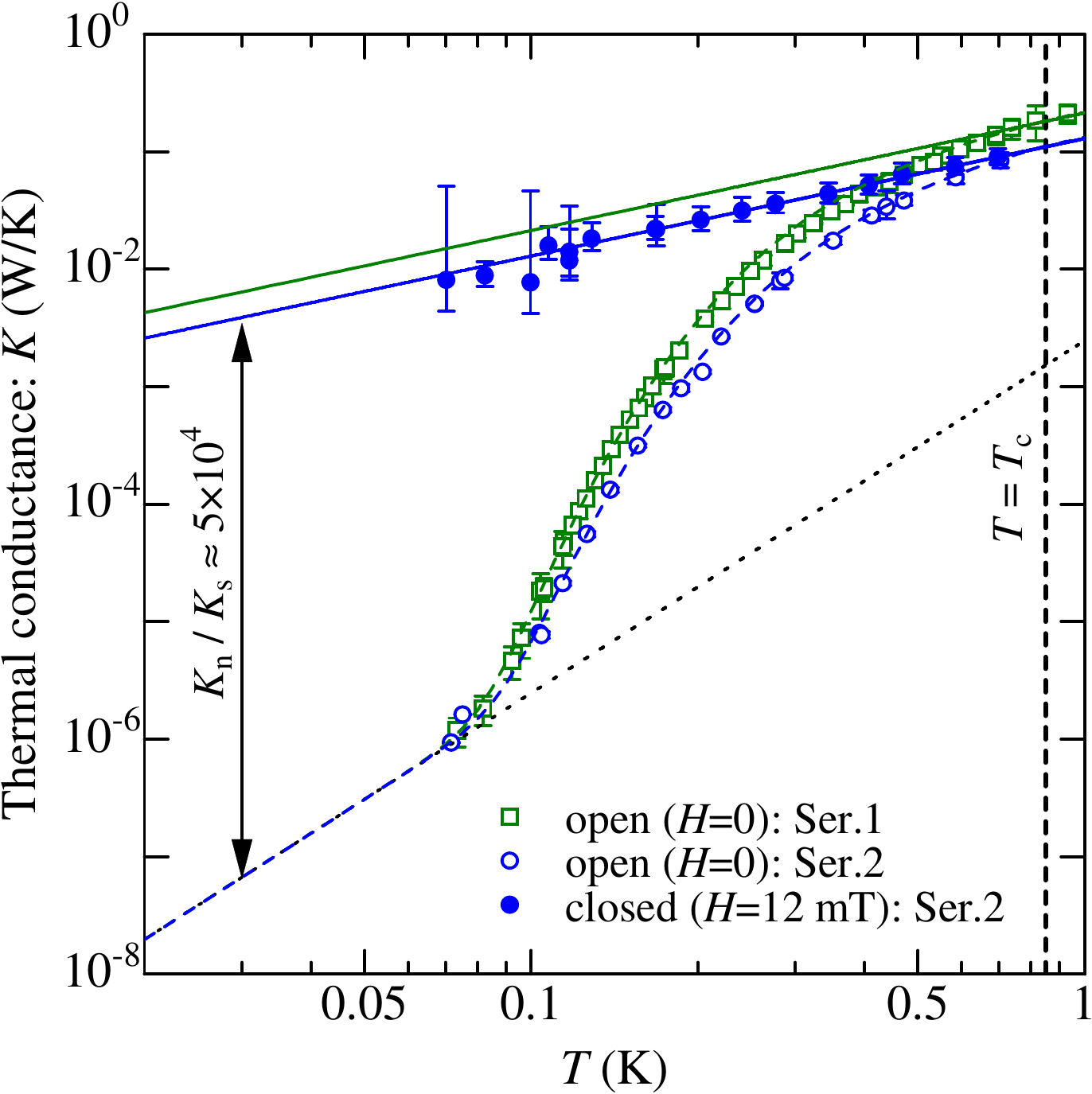}
   \caption{Open-mode thermal conductance of the HSW measured in the first series ($K{_\mathrm{open1}}$; open squares), closed-mode ($K{_\mathrm{closed2}}$; solid circles) and open-mode ($K{_\mathrm{open2}}$; open circles) thermal conductances of the assembled HSW measured in the second series. The solid lines are fittings of the $K{_\mathrm{open1}}$ data at $T \geq T_\mathrm{c}$ and the $K{_\mathrm{closed2}}$ data to $T$-linear functions. The dashed lines are numerical calculations for the $K{_\mathrm{open1}}$ and $K{_\mathrm{open2}}$ data Eq.(\ref{eq: BCS}) expected from the BCS theory. The dotted line is the  phonon contribution to $K{_\mathrm{open2}}$ expected from Eq.(\ref{eq: eq1}) with $\lambda_\mathrm{ph} = 0.25$~mm.}
\label{fig:conductance}
\end{figure}

Results of the first measurement of the thermal conductance in the open mode $K{_\mathrm{open1}}$ without applying external magnetic field ($H$) are shown as (green) open squares in Fig.~\ref{fig:conductance}. 
In this measurement series (Ser.1), because the magnet for switching was not installed, we did not measure the thermal conductance in the closed mode.
The $K{_\mathrm{open1}}$ data follow very well the (green) dashed line which is a calculation based on the BCS theory~\cite{Peshkov1965}:
\begin{equation}
\label{eq: BCS}
 K_\mathrm{el,s} = K_\mathrm{el,s}(T_\mathrm{c})\exp \left[-\frac{\Delta}{k_\mathrm{B} T}\left(1-\frac{T}{T_\mathrm{c}}\right)\right]
                 = \frac{L T_\mathrm{c}}{R_\mathrm{s}} \exp\left[-\frac{\Delta}{k_\mathrm{B} T}\left(1-\frac{T}{T_\mathrm{c}}\right)\right] 
\end{equation}
where $T_\mathrm{c} =$~850~mK. The $R_\mathrm{s}$ is residual electrical resistance of zinc region and contact region whose thermal conductances are affected by the superconducting transition. Because the silver region remains normal state, thermal conductivity of this region should be proportional to the temperature and inversely proportional to the residual electrical resistance of silver region ($R_\mathrm{n}$). In the calculation, we take into account these series conductance by electron and conductance by phonons in zinc part. For the Ser.1, we use the energy gap $\Delta =$~1.4$k_\mathrm{B} T_\mathrm{c}$ and $R_\mathrm{s} =$~59~n$\mathrm{\Omega}$ and $R_\mathrm{n} =$~55~n$\mathrm{\Omega}$ to fit measured value. The $R_\mathrm{s}$ is larger than expected value (about 41~$\mathrm{\Omega}$), and $R_\mathrm{n}$ is smaller than expected value (about 66~n$\mathrm{\Omega}$). This $\Delta$ value is consistent with in the range of previous reported value for zinc\cite{Zavaritskii1961, Contignola1967}.
At the lowest temperature ($= 70$~mK), the data approach the (black) dotted line which is a phonon contribution expected from Eq.(\ref{eq: eq1}) with $\lambda_\mathrm{ph} = 0.25$~mm. 
From the data points of $K{_\mathrm{open1}}$ at $T \geq T_\mathrm{c}$, we can  estimate the thermal conductance in the closed mode as $K{_\mathrm{closed1}}$~(W/K) $= (0.22\pm 0.03) T$ (the (green) solid line).
This result is consistent within the estimation error (1\%) with the  $R_\mathrm{HSW}$ value measured just before the thermal conductance measurement, indicating the validity of the Wiedemann-Franz law with the standard Lorenz number $L$  for Zn.
For Al and Ag, the anomalously small Lorentz numbers are reported  in the previous work~\cite{Gloos1990} where a factor of 2--20 reduction of $L$ is claimed from the measurements at $2 \leq T \leq 8$~K.
However, we did not observe such an anomalous behavior at least for the Zn/Ag contact foils in the temperature range of $70 \leq T \leq 900$~mK (combined with the result shown just below).

The thermal conductance in the closed mode $K{_\mathrm{closed2}}$ was measured down to 70~mK in the subsequent measurement series (Ser.2) in the applied field of $H = 12$~mT .
The result  is shown as the (blue) closed circles, which follow nicely the expected $T$-linear behavior: $K{_\mathrm{closed2}}$~(W/K) $= 0.13 T$ (the (blue) solid line).
The coefficient is smaller than that of $K{_\mathrm{closed1}}$ by a factor of 1.7 because, in between the two measurement series, a large mechanical stress had been applied to the Zn foil due to accidental bending and straightening of the assembly by 8~degrees.
After this experience, we added thin glassfiber support plates to protect the  soft Zn foils in actual experiments. 
Concomitantly, the measured $K{_\mathrm{open2}}$ data, which are shown as the (blue) open circles, are smaller than the $K{_\mathrm{open1}}$ data by the same factor of 1.7 near $T = T{_\mathrm{c}}$, but,
near the lowest temperature ($= 70$~mK), they approach the same (black) dotted line as that for $K{_\mathrm{open1}}$ to lower temperature, indicating that the mechanical damage affected only on $\lambda{_\mathrm{el}}$ not on $\lambda{_\mathrm{ph}}$.
For the Ser.2, we use the same $\Delta =$~1.4$k_\mathrm{B} T_\mathrm{c}$ and $R_\mathrm{n} =$~55~n$\mathrm{\Omega}$ and $R_\mathrm{s} =$~133~n$\mathrm{\Omega}$ to fit the experimental data. The $R_\mathrm{s}$ get worse than before.

By extrapolating the measured $K{_\mathrm{closed2}}$ and $K{_\mathrm{open2}}$ data, we can safely estimate the switching ratio $K{_\mathrm{closed}}/K{_\mathrm{open}}$ at 30~mK and 0.8~mK as $5 \times 10^4$ and $7 \times 10^7$, respectively.


\section{Summary}
\label{sec:conc}
A compact and high-performance superconducting heat switch (HSW) of zinc (Zn) suited for the recently proposed continuous nuclear demagnetization refrigerator (CNDR) was developed.
The CNDR, which can maintain $T = 0.8$~mK continuously under a heat load of 10~nW, consists of serially connected two PrNi$_5$ nuclear stages and two HSWs.

By considering requirements for this application carefully and by optimizing various heat treatments of construction materials, we determined a detailed design recipe for HSWs usable at sub-mK.
Following the recipe, we constructed an HSW for the CNDR with six pure Zn foils of 0.25~mm thick and seven pure silver (Ag) foils of 0.5~mm thick as a thermal link.
During the course of the design and preparation, we found an optimum temperature ($= 220$~$^\circ$C) to exist for heat treatment to diffusively bond Zn foil to Ag one.

The residual electrical resistance measured at 4.2~K ($\rho_{\mathrm{HSW}}$) of the whole HSW assembly was 114~n$\mathrm{\Omega}$, which is consistent with the estimation from the measured resistivities of various components.
The measured thermal conductance in the closed mode follows nicely a $T$-linear behavior down to 70~mK, and the proportion coefficient agrees very well with the expectation from the $\rho_{\mathrm{HSW}}$ value.
In other words, we confirmed the validity of the Wiedemann-Franz law with the normal Lorenz number for our Zn/Ag foils.
Since this result is different from anomalously low Lorenz numbers for pure aluminum and silver  reported previously by other workers at higher temperatures, more comprehensive studies will be necessary to address such an apparent discrepancy.
The thermal conductance in the open mode was also measured in a temperature range of $70 \leq T \leq 900$~mK, being consistent with the prediction of the BCS theory.
The result indicates that the heat leak through this HSW in the open mode will be 0.5~nW at most and practically less than 0.2~nW.

Once the CNDR becomes available, it will expand utility of the sub-mK environment to broader research fields.
Actual performances of this HSW as a part of the CNDR will be tested in the near future.
The HSW constructed in this work and the design recipe described here are conveniently applicable to heat capacity measurements at sub-mK temperatures as well.

\section{Acknowledgements}
This work was financially supported by Grant-in-Aid for Challenging Exploratory Research (Grant No.~15K13398) from JSPS.

\bibliography{hswbibfile}

\end{document}